\documentclass[twocolumn,preprintnumbers, amssymb,amsmath,aps,floatfix,prd,nofootinbib,superscriptaddress,showpacs]{revtex4-2}

\usepackage{epsfig}
\usepackage{bm}
\usepackage{amssymb}
\usepackage{amsmath}
\usepackage{comment}
\usepackage{slashed}
\usepackage{simpler-wick}
\usepackage{color}
\usepackage{subfigure}
\usepackage[colorlinks,
            linkcolor=blue,
            anchorcolor=black,
            citecolor=blue
            ]{hyperref}

\newcommand{\beq}{\begin{eqnarray}}
\newcommand{\eeq}{\end{eqnarray}}

\begin{document}
\title{Probing gluonic saturation in deeply virtual meson production beyond leading power
}

\author{Renaud Boussarie}
\email{Renaud.Boussarie@polytechnique.edu}
\affiliation{CPHT, CNRS, Ecole Polytechnique, Institut Polytechnique de Paris, 91128 Palaiseau, France}

\author{Michael Fucilla}
\email{Michael.Fucilla@ijclab.in2p3.fr}
\affiliation{Université Paris-Saclay, CNRS/IN2P3, IJCLab, 91405, Orsay, France}

\author{Lech Szymanowski}
\email{Lech.Szymanowski@ncbj.gov.pl}
\affiliation{National Centre for Nuclear Research (NCBJ), Pasteura 7, 02-093 Warsaw,  Poland}

\author{Samuel Wallon}
\email{Samuel.Wallon@ijclab.in2p3.fr}
\affiliation{Université Paris-Saclay, CNRS/IN2P3, IJCLab, 91405, Orsay, France}

\begin{abstract}
Exclusive diffractive meson production represents a golden channel for investigating gluonic saturation inside nucleons and nuclei. In this letter, we settle a systematic framework to deal with beyond leading power corrections at small-$x$, including the saturation regime, and obtain the $\gamma^{*} \rightarrow M (\rho, \phi, \omega)$ impact factor with both incoming photon and outgoing meson carrying arbitrary polarizations. This is of particular interest since the saturation scale at modern colliders, although entering a perturbative regime, is not large enough to prevents higher-twist effects to be sizable.
\end{abstract}
\maketitle

\section*{Introduction}

Gluonic saturation within nucleons and nuclei represents one of the most fascinating aspects of Quantum Chromodynamics and one of the pillars of the Electron-Ion-Collider (EIC) physics program~\cite{AbdulKhalek:2021gbh}. Multiple efforts are needed in order to produce theoretical predictions that can unambiguously reveal the state of hadronic matter that is formed in the saturation regime, the Color Glass Condensate (CGC). A robust description of the hadronic matter in the high-energy and dense regime of QCD is provided by the semi-classical McLerran-Venugopalan (MV) model~\cite{McLerran:1993ka,McLerran:1993ni,McLerran:1994vd}. Quantum corrections in the CGC lead to the well-known Balitsky---Jalilian Marian-Iancu-McLerran-Weigert-Leonidov-Kovner (B-JIMWLK)~\cite{Balitsky:1995ub,Jalilian-Marian:1997qno,Jalilian-Marian:1997ubg,Jalilian-Marian:1997jhx,Jalilian-Marian:1998tzv,Weigert:2000gi,Iancu:2000hn,Iancu:2001ad} evolution equations. The MV model, supported by a leading order (LO) B-JIMWLK evolution, provides an excellent qualitative description of the saturation phenomena. Nonetheless, to unambiguously reveal and study the nature of the CGC, the precision physics frontier at small $x$ must be achieved. To this aim, large corrections must be taken into account in order to compare experimental measurements from modern and future colliders. Among them, there are the one in powers of $\alpha_s$, i.e. next-to-leading order (NLO) corrections~\cite{Kovchegov:2006vj,Balitsky:2007feb,Balitsky:2013fea,Grabovsky:2013mba,Balitsky:2014mca,Kovner:2013ona,Lublinsky:2016meo,Balitsky:2010ze,Balitsky:2012bs,Chirilli:2011km,Iancu:2016vyg,Boussarie:2014lxa,Boussarie:2016ogo,Fucilla:2022wcg,Fucilla:2023mkl,Beuf:2022ndu,Altinoluk:2020qet,Taels:2022tza,Roy:2019hwr,Caucal:2021ent,Caucal:2022ulg,Caucal:2023fsf,Bergabo:2022zhe,Bergabo:2022tcu,Iancu:2022gpw,Mantysaari:2021ryb,Mantysaari:2022kdm,Beuf:2022kyp,Taels:2023czt,Caucal:2024cdq}, and the one in the inverse power of $\sqrt{s}$, known as subeikonal contributions~\cite{Altinoluk:2014oxa,Chirilli:2018kkw}.
A further important aspect is that many observables can be sensitive to hard scale power-suppressed contributions, known as higher-twist effects. In fact, saturation effects are expected to occur at hard scales of the order of the saturation scale, which at the EIC will not be large enough to justify a strong suppression of higher-twist effects. 
It is therefore essential to include higher-twist contributions in precision studies. The aim of this paper is to develop a general method to accommodate this need, which we illustrate in the paradigmatic example of the electroproduction of a light vector meson. \\

The electroproduction of a light vector meson has been extensively studied at the HERA experiments~\cite{ZEUS:2007iet,Aaron:2009xp,ZEUS:1999ptu,ZEUS:2002vvv,H1:2006ogl}. 
These experiments showed a theoretically unexpected largeness of the contribution from transversaly polarized mesons, which requires the computation of power suppressed contributions. First, the total cross section receives contribution from both longitudinally and transversely polarized produced mesons, with these latter suppressed by a relative factor $\left( m_M / Q \right)^2$, where $m_M$ is the mass of the meson and $Q$ is the virtuality of the incoming virtual photon. Even considering the largest saturation scale, $Q_s^2 \sim 2.2$ GeV$^2$, reachable at the EIC in DIS off heavy ion~\cite{Aschenauer:2017jsk}, and a squared photon virtuality of almost the same order, the contribution of the transversally polarized meson to the total cross section is theoretically expected to be of order $10\%$. However, measurements carried out at different experiments show that the effect appears to be much more important then expected. In particular, the $Q^2$ dependence of the ratio $R= \sigma_L / \sigma_T$ indicates that $\sigma_L$ becomes dominant only for $Q^2$ larger than 2 GeV$^2$~\cite{COMPASS:2022xig}. Moreover, even at higher values of photon virtuality where $\sigma_L$ dominates, e.g. $ 2 < Q^2 < 10$ GeV$^2$, the $\sigma_T$ contribution is very sizable~\cite{COMPASS:2022xig,Aaron:2009xp}. Therefore, in the kinematic region sensitive to saturation effects, the longitudinal-to-longitudinal and transverse-to-transverse contributions are comparable. Beside that, a strong evidence of gluonic saturation obviously requires analysing the largest possible set of data. Besides these total cross section measurements, the H1~\cite{H1:2000hps,Aaron:2009xp} and ZEUS~\cite{ZEUS:2005bhf} experiments have provided an exhaustive analysis of the spin-density matrix elements of the $\gamma^{*} (\lambda_{\gamma}) P \rightarrow M (\lambda_{M}) P$ process, the largest of which is the one describing the longitudinal photon to longitudinal meson transition.
This latter, from a theoretical point of view, corresponds to the leading twist contribution and the description of all other transitions requires a treatment beyond leading twist. The above mentioned observables are of special interest at the future EIC experiments. \\

Thus, in this letter, we consider deeply virtual meson production (DVMP) in the high-energy limit, describing, for the first time, all helicity amplitudes, including the ones which are suppressed by one power of the photon virtuality and require a systematic treatment beyond the leading twist. We perform the calculation in the non-forward kinematics, in order to be able to probe the generalized transverse momentum dependent (GTMD) gluon distribution of the target. Moreover, we describe for the first time the transversally polarized light vector-meson production, which starts at the next-to-leading power, in the saturation regime. The new operatorial method we are developing here permits the systematic inclusion of genuine higher-twist effects, coming from higher Fock state wavefunctions, i.e. with a non-minimal parton configuration in the vector meson. \\

Finally, the presently developed method, which is systematic and universal, could be directly used in order to obtain power suppressed contributions to processes which, although being twist-2 dominated, would presumably require higher-power contributions to be taken into account to unveil interesting physical effects and for precision physics. A suggestive example is the exclusive photoproduction of a photon-pion pair, $\gamma + P \rightarrow \gamma + \pi^0 + P$, in which the invariant mass, $M_{\pi^0 \gamma}$, of the outgoing $\pi^0 \gamma$ system plays the role of the hard scale. This process, similarly to the processes investigated in this paper, violates collinear factorization, but with the breaking happening already at the leading power in the $M_{\pi^0 \gamma}$ expansion~\cite{Nabeebaccus:2023rzr,Nabeebaccus:2024mia}. The leading twist result can be easily obtained by using the framework of the paper, to build a finite result in $k_T$-factorization. Nonetheless, the process is of particular interest beyond the leading twist. Indeed, it then becomes sensitive to the magnetic susceptibility of the quark condensate~\cite{Ball:2002ps}, which probes the hadronic content of the photon. A similar sensitivity to the magnetic susceptibility appears, beyond the leading twist, in the exclusive pion photoproduction, $\gamma + P \rightarrow \pi^0 + P$, at large momentum transferred squared\footnote{Here the momentum transferred in the $t$-channel squared plays the role of the hard scale.}. In this channel one can expect higher twist contaminations to be large, in analogy to the electroproduction channel where higher-twist corrections are essential to describe medium energy JLab data~\cite{JeffersonLabHallA:2016wye}.

\section*{$k_T$-factorization and semi-classical approach to saturation}

We rely on the $k_T$-factorization and introduce a light-cone basis using vectors $n_1$ and $n_2$, which specify the $+/-$ directions and such that $n_1 \cdot n_2 = 1$. 
The Sudakov decomposition of any vector $k$ is expressed as\footnote{Transverse momentum in Euclidean space are denoted in bold characters, while Minkowskian ones by the $\perp$ subscript.} $k^\mu = k^+ n_1^\mu + k^- n_2^\mu + k_\perp^\mu $ and $q$ ($p$) denotes the incoming photon (proton) four-momentum. We consider the semi-hard kinematics $s \gg Q^2 = - q^2 \gg \Lambda_{\text{QCD}}^2$, where $s$ is the center-of-mass energy of the photon-proton system and $\Lambda_{\text{QCD}}$ is the QCD scale parameter. In the following, we will choose a reference frame, referred to as the projectile frame, where the target moves ultra-relativistically towards the probe. The projectile travels along the $n_1$ (or $+$) direction, while the target has a large component along the $n_2$ (or $-$) direction, which also defines our QCD gauge choice ($A \cdot n_2 = 0$). \\

In the present framework, we adopt the semi-classical~\cite{McLerran:1993ka,McLerran:1993ni,McLerran:1994vd} effective approach to QCD at small $x$~\cite{Balitsky:1995ub,Balitsky:1998kc,Balitsky:1998ya,Balitsky:2001re}, in which the gluonic field $\mathcal{A}(k)$ is split into classical background fields $b(k)$ (small $k^+$) and quantum fields $A(k)$ (large $k^+$).    
In the high-energy limit, one can perform a very large boost from the target rest frame to the projectile frame, so that the external field can be written as
\begin{equation}
    b^\mu (x) = b^-(x^+, x_\perp) n_2^\mu = \delta (x^+) \mathbf{B} (\boldsymbol{x})  n_2^\mu \,.
\end{equation}
This is known as the shockwave approximation.  \\

In this approach, resummation of all-order interactions with the background field yields an infinite null Wilson line located precisely at $z^- = 0$, 
\begin{equation}
   W_{\boldsymbol{z}} = \mathcal{P} \exp \left(i g \int d z^+ b^{-a}(z) \mathbf{T}^a_r \right)\, ,
   \label{Eq:GenericWilsonLine}
\end{equation}
where $\mathcal{P}$ denotes the path ordering operator in the $+$ direction and $\mathbf{T}^a_r$ is a color matrix in the $r$ representation of the gauge group. In the following, we use $V_{\boldsymbol{z}}$ and $U_{\boldsymbol{z}}$ to denote Wilson lines in the fundamental ($\mathbf{T}^a_r \rightarrow t^a_{ij}$) and adjoint ($\mathbf{T}^a_r \rightarrow T^a_{bc}$) representations respectively. Using small-$x$ factorization, the scattering amplitude is expressed as the convolution of the projectile impact factor with the non-perturbative matrix element of Wilson line operators between target states. \\

Eikonal interactions between the projectile and the classical fields can be resummed into effective operators ($\psi_{\text{eff}}$, $\bar{\psi}_{\text{eff }}$, $A_{\text {eff }}^{\mu a}$)~\cite{Boussarie:2024bdo}. For the fermionic field, we get
\begin{equation}
    \left[\psi_{\text {eff }}\left(z_0\right)\right]_{z_0^{+}<0} = - \int \mathrm{d}^D z_2 G_0\left(z_{02}\right) V_{\boldsymbol{z}_2}^{\dagger} \gamma^{+} \psi\left(z_2\right) \delta\left(z_2^{+}\right) 
    \label{Eq:PsiEffecNoMon}
\end{equation}
and
\begin{equation}
    \left[\bar{\psi}_{\text{eff }}\left(z_0\right)\right]_{z_0^{+}<0} = \int \mathrm{d}^D z_1 \bar{\psi}\left(z_1\right) \gamma^{+} V_{\boldsymbol{z}_1} G_0\left(z_{10}\right) \delta\left(z_1^{+}\right) \; ,
    \label{Eq:PsiBarEffecNoMon}
\end{equation}
where $G_0(x)$ denotes the free quark propagator, 
\begin{equation}
    G_0 (z) = \int \frac{d^D l}{(2 \pi)^D} e^{-i l \cdot z} \frac{i \slashed{k}}{k^2+i0} \; .
\end{equation}
The eq.~(\ref{Eq:PsiEffecNoMon}) (eq.~(\ref{Eq:PsiBarEffecNoMon})) represents a fermionic line starting at the light-cone time $z_0^+ < 0$ and freely propagating to $z_2^+$ ($z_1^+$) where it eikonally interacts with the background shockwave field. The motivation for introducing such operators is that they serve to construct amplitudes involving non-perturbative matrix elements of general off light-cone correlators, i.e. without any reference to the twist-expansion. \\

\noindent In the same spirit, the gluonic effective operator reads
\begin{equation}
   \left[ A_{\text {eff }}^{\mu a}\left(z_0\right) \right]_{z_0^{+}<0} \hspace{-0.1 cm } = 2 i \hspace{-0.1 cm } \int \hspace{-0.1 cm } \mathrm{d}^D z_3 \delta\left(z_3^{+}\right) F_{-\sigma}^b\left(z_3\right)   G^{\mu \sigma_{\perp}} \hspace{-0.1 cm } \left( z_{30} \right) U_{\boldsymbol{z}_3}^{a b} \; ,
   \label{Eq:AEffecNoMon}
\end{equation}
where $G^{\mu \sigma_{\perp}} \left( x \right)$ is the free gluon propagator in the $n_2$ light-cone gauge,
\begin{equation}
    G^{\mu \sigma_{\perp}} (z) = \int \frac{d^D l}{(2 \pi)^D} e^{-i l \cdot z} \frac{-i}{l^2+i0} \left( g^{\mu \sigma}_{ \perp} - \frac{n_2^{\mu} l_{\perp}^{\sigma}}{l^+} \right)  \; ,
\end{equation}
and $F_{\mu \nu}(x) = \partial_{\mu} A_{\nu}^a (z) - \partial_{\nu} A_{\mu}^a (z) - g f^{abc} A_{\mu}^b (z) A_{\nu}^{b} (z)$ is the QCD field strength tensor.

\section*{Exclusive light vector meson production}
A complete twist-3 computation of the DVMP processes requires to correctly account for both kinematic and genuine twist effects. These two effects are physically different, but related to each other by the QCD equations of motion and, therefore, must be taken into account simultaneously. \\

\begin{figure}
\begin{center}
\includegraphics[width=0.2\textwidth]{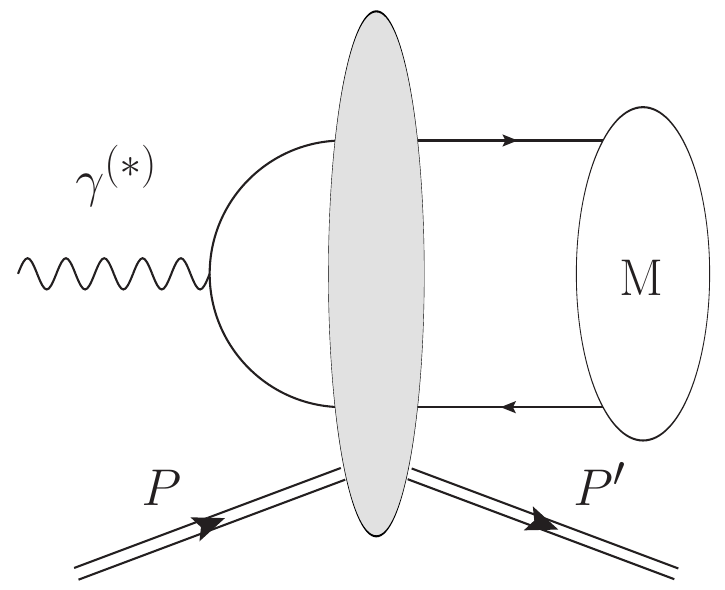} 
\end{center}
\caption[*]{2-body diagram contributing to the $\gamma^{*} (\lambda_{\gamma}) P \rightarrow M (\lambda_{M}) P$ process within twist-3 accuracy}
\label{fig:2-body}
\end{figure}

\textit{2-body contribution}. The kinematic twist effects are related to the dependence of the leading Fock state wave function (i.e. with a minimal number of valence partons) on the relative transverse momentum of the constituents. In our case, this corresponds to calculate the diagram in fig.~\ref{fig:2-body}, allowing for a relative transverse momentum (or equivalently a space-like separation) between the quark and the antiquark that form the meson. Using the effective operators in eq.~(\ref{Eq:PsiEffecNoMon}) and (\ref{Eq:PsiBarEffecNoMon}), the contribution in fig.~\ref{fig:2-body} is written as
\begin{gather}
{\cal A}_2 =-ie_{q}\int{\rm d}^{D}z_{0}\theta\left(-z_{0}^{+}\right){\rm e}^{-i\left(q\cdot z_{0}\right)} \nonumber \\ \times \left\langle P\left(p^{\prime}\right)M\left(p_{M}\right)\left|\overline{\psi}_{{\rm eff}}\left(z_{0}\right) \slashed{\varepsilon}_{q}\psi_{{\rm eff}}\left(z_{0}\right)\right|P\left(p\right)\right \rangle \; , 
\label{Eq:GeneralStructA2}
\end{gather}
where $\varepsilon_{q}$ is the photon polarization vector, $e_q$ the electromagnetic charge of the quark, $p_M$ ($p'$) the four-momentum of the outgoing meson (proton). After subtraction of the non-interacting part\footnote{The contribution with all Wilson lines set to the identity.}, the amplitude is factorized into a matrix element between the proton states of the dipole operator,
\begin{gather}
    \mathcal{U}_{\boldsymbol{b}+\overline{x}\boldsymbol{r} \; \boldsymbol{b}-x\boldsymbol{r}} = 
    1-\frac{1}{N_{c}}{\rm tr}\left(V_{\boldsymbol{b}+\overline{x}\boldsymbol{r}}V_{\boldsymbol{b}-x\boldsymbol{r}}^{\dagger}\right) \; ,
\end{gather}
and a photon/meson wave functions overlap (PMWO) $\Psi_2 \left(x, \boldsymbol{r} \right)$. The amplitude reads 
\begin{gather}
   \mathcal{A}_2 = \int_{0}^{1} {\rm d} x \int{\rm d}^{2} \boldsymbol{r} \Psi_2 \left(x, \boldsymbol{r} \right) \int{\rm d}^{d} \boldsymbol{b} \; {\rm e}^{i (\boldsymbol{q}-\boldsymbol{p}_M) \cdot\boldsymbol{b}} \nonumber \\ \times \left\langle P\left(p^{\prime}\right)\left|\mathcal{U}_{\boldsymbol{b}+\overline{x}\boldsymbol{r} \; \boldsymbol{b}-x\boldsymbol{r}} \right|P\left(p\right)\right\rangle \; ,
\label{Eq:StandardDipoleAmp-rb}
\end{gather}
where we introduced the standard variables (momentum fraction, dipole size and impact parameter) $x,\boldsymbol{r}$ and $\boldsymbol{b}$ and $\bar{x} = 1-x$. \\

Without any further assumption, a tedious but straightforward computation~\cite{Boussarie:2024bdo} gives the PMWO
\begin{widetext} 
\begin{gather}
    \Psi_2 = e_q \delta \left( 1 - \frac{p_M^+}{q^+} \right) \left(\varepsilon_{q\mu}-\frac{\varepsilon_{q}^{+}}{q^{+}}q_{\mu} \right) 
  \left[ \phi_{\gamma^+} \left( 2 x \bar{x} q^{\mu} - i (x-\bar{x}) \frac{\partial}{\partial r_{\perp \mu}} \right) + \epsilon^{\mu\nu+-} \phi_{\gamma^+ \gamma^5}  \frac{\partial}{\partial r_{\perp}^{\nu}} \right] K_0 \left( \sqrt{x \bar{x} Q^2 \boldsymbol{r}^{2}} \right) \; ,
  \label{Eq:Psi_2_Full_KinTwist}
\end{gather}
\end{widetext}
with 
\begin{equation}
\phi_{\Gamma^{\lambda}} \hspace{-0.1 cm} = \hspace{-0.2 cm} \int  \hspace{-0.1 cm} \frac{ d r^-}{2 \pi} e^{i x \boldsymbol{p}_M \cdot\boldsymbol{r} -i x p_M^+ r^- } \langle M(p_{M})\left| \overline{\psi}\left(r\right)\Gamma^{\lambda}\psi\left(0\right)\right|0\rangle_{r^+=0}     \label{Phi_+_Distrib_Coordinate}
\end{equation}
where $\Gamma^{\lambda}= \{ \gamma^+, \gamma^+ \gamma^5 \}$.
We emphasize here that the PMWO in~(\ref{Eq:Psi_2_Full_KinTwist}) still contains all-twist kinematic effects. Indeed, the vacuum-to-meson matrix elements of nonlocal operators, in eq.~(\ref{Phi_+_Distrib_Coordinate}), are off light-cone due to the non-vanishing $r_{\perp}$ component (while $r^+=0$ due to the shockwave approximation), so that the quark and the antiquark are not collinear to each other. The term proportional to the momentum $q^{\mu}$ in eq.~(\ref{Eq:Psi_2_Full_KinTwist}) contains both the longitudinal photon to longitudinal meson transition, starting at twist 2, and the longitudinal photon to transverse meson transition, starting at twist 3. The twist-2 contribution is already known at the NLO~\cite{Ivanov:2004pp,Boussarie:2016bkq,Mantysaari:2022bsp} and we thus omit it in the following discussion. The remaining terms are related to the transverse photon to transverse meson transition and contribute to the amplitude starting from the twist-3 term. Non-forward transitions such as L-to-T and T-to-L have been computed in a saturation context for the present~\cite{Boussarie:2016bkq} and other processes~\cite{Kowalski:2006hc,Hatta:2017cte,Mantysaari:2020lhf}, and involve similar 2-body wave function overlaps. We stress that the 3-body overlaps were not included in those computations despite being of the same order in twist. \\

\begin{figure}
\begin{center}
\includegraphics[width=0.2\textwidth]{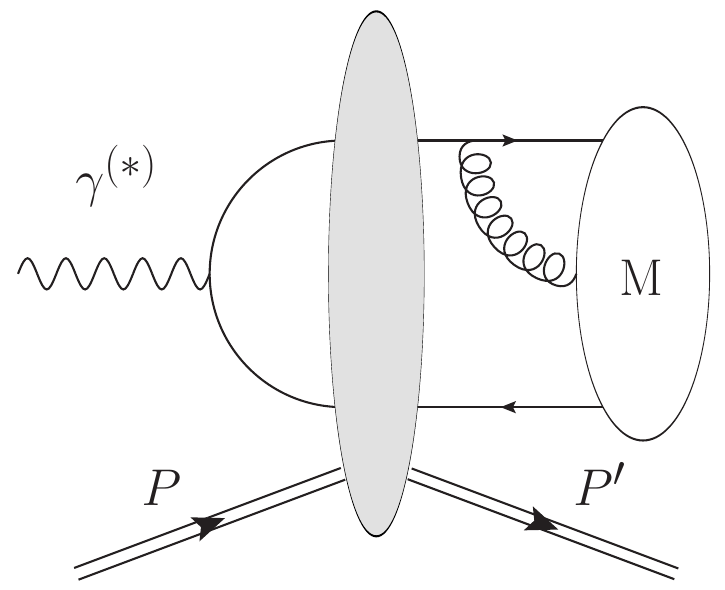} \hspace{0.5 cm}
\includegraphics[width=0.2\textwidth]{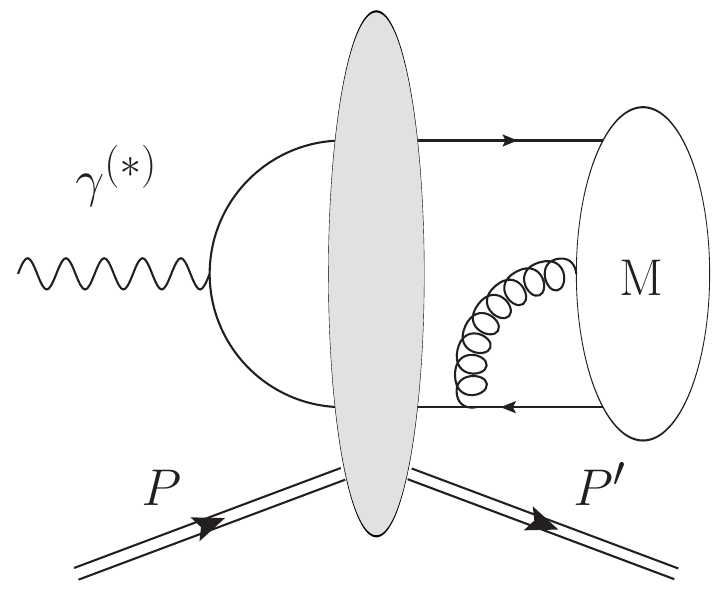}
\end{center}
\begin{center}
\includegraphics[width=0.2\textwidth]{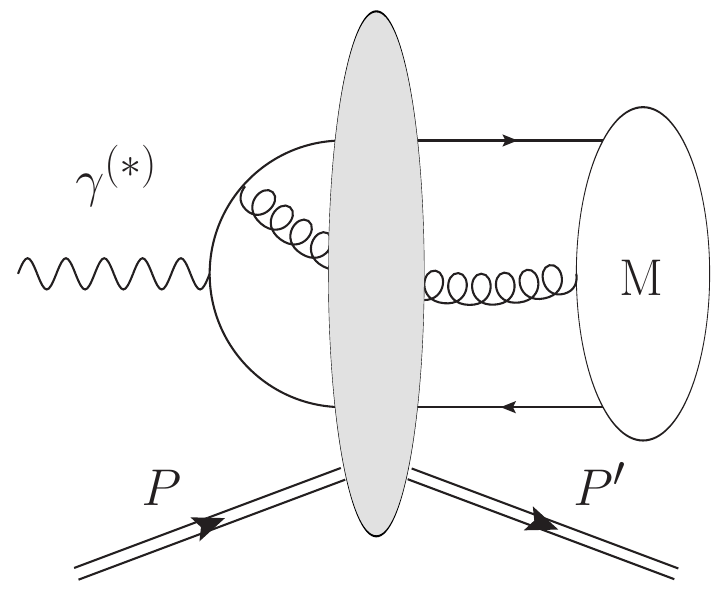} \hspace{0.5 cm}
\includegraphics[width=0.2\textwidth]{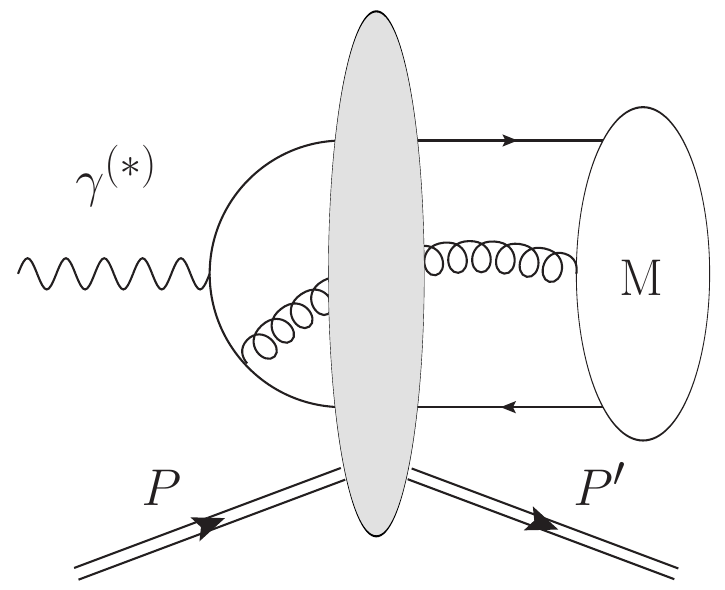}
\end{center}
\caption[*]{3-body diagrams contributing to the $\gamma^{*} (\lambda_{\gamma}) P \rightarrow M (\lambda_{M}) P$ process within twist-3 accuracy}
\label{fig:3-body}
\end{figure}

\textit{3-body contribution}. The genuine twist 3 effects are related to the appearance, at the same perturbative order, of higher Fock states (i.e. with a non-minimal parton configuration). In our case, these contributions are related to diagrams with an additional emitted gluon, pictorially depicted in fig.~\ref{fig:3-body}. The top diagrams in fig.~\ref{fig:3-body}, i.e. those in which the gluon is emitted after the shockwave, do not contribute to our coefficient function (the impact factor) within twist 3 accuracy. In fact, the twist-3 contribution of diagrams in fig.~\ref{fig:3-body} is obtained by requiring the three partons that connect the hard and soft part of the process to fly collinearly. Then, in the top diagrams, the quark (left diagram) or antiquark (right diagram) propagator after the shockwave automatically goes on its mass-shell. This implies that these contributions become purely singular and are therefore associated with the Efremov-Radyushkin-Brodsky-Lepage (ERBL) evolution of the twist-2 DAs~\cite{Farrar:1979aw,Lepage:1979zb,Efremov:1979qk}. The diagrams at the bottom of fig.~\ref{fig:3-body}, can be calculated similarly to the contribution in eq.~(\ref{Eq:GeneralStructA2}); for instance, the contribution in which the gluon is emitted from the quark line reads\footnote{The contribution in which the gluon is emitted from the anti-quark is similarly obtained.}
\begin{widetext}
\begin{equation}
{\cal A}_{q} =e_{q}g\int{\rm d}^{D}z_{4}{\rm d}^{D}z_{0}\theta\left(-z_{4}^{+}\right)\theta\left(-z_{0}^{+}\right) \left\langle P\left(p^{\prime}\right)M\left(p_{M}\right)\left|\overline{\psi}_{{\rm eff}}\left(z_{4}\right)\gamma_{\mu}A_{{\rm eff}}^{\mu a}\left(z_{4}\right)t^{a}G\left(z_{40}\right)\slashed{\varepsilon}_{q}{\rm e}^{-i\left(q\cdot z_{0}\right)}\psi_{{\rm eff}}\left(z_{0}\right)\right|P\left(p\right)\right\rangle \; .
\end{equation}
After relevant projections in Dirac and color space, we can again separate the dynamics of target and projectile. The crossing of the shockwave by the gluon induces the appearance of a double dipole operator $\mathcal{U}_{\boldsymbol{z}_1 \boldsymbol{z}_3} \mathcal{U}_{\boldsymbol{z}_3 \boldsymbol{z}_2}$. Then, once the non-interacting contribution is subtracted, we can again factorize the amplitude into a 3-body PMWO and a matrix element between the proton states, i.e. 
    \begin{gather}
    \mathcal{A}_3 = \left( \prod_{i=1}^3 \int \hspace{-0.1 cm} d x_i \hspace{-0.1 cm} \int \hspace{-0.1 cm} d^2 \boldsymbol{z}_i e^{i x_i \boldsymbol{q} \boldsymbol{z}_i} \theta (x_i) \right) \hspace{-0.1 cm} \delta (1 - \sum_i x_i )
    \Psi_3 \left( \{ x_i \} , \{ \boldsymbol{z}_i \} \right) \left\langle P\left(p^{\prime}\right)\left| \mathcal{U}_{\boldsymbol{z}_1 \boldsymbol{z}_3} \mathcal{U}_{\boldsymbol{z}_3 \boldsymbol{z}_2} - \mathcal{U}_{\boldsymbol{z}_1 \boldsymbol{z}_3} -\mathcal{U}_{\boldsymbol{z}_3 \boldsymbol{z}_2} + \frac{ \mathcal{U}_{\boldsymbol{z}_1 \boldsymbol{z}_2}}{N_c^2}  \right|P\left(p\right)\right\rangle ,
   \label{Eq:GenealStructure3bodyWaveFunOver}
\end{gather}
where $\{ a_i \} = \{ a_1, a_2, a_3 \}$.  \\

\noindent We calculate separately the contribution from quark- and antiquark-gluon emission in fig.~\ref{fig:3-body}, separating them into a QED gauge invariant and a QED gauge breaking term. The sum of the two QED gauge breaking terms vanishes leading to the explicitly QED gauge invariant 3-body PMWO~\cite{Boussarie:2024bdo} 
\begin{gather}
    \Psi_3 \left( \{ x_i \} , \{ \boldsymbol{z}_i \} \right) = \frac{e_{q} q^{+}}{8 \pi}  \left(\varepsilon_{q\rho}-\frac{\varepsilon_{q}^{+}}{q^{+}}q_{\rho}\right) c_f \bigg \{ \chi_{\gamma^+ \sigma } \left[ \left( 4 i g_{\perp \perp}^{ \rho \sigma } \frac{x_1 x_2}{1-x_2} \frac{Q}{Z} K_1 (QZ) +  T_1^{\sigma \rho \nu} (\{ x_i \}) \frac{z_{23 \perp \nu}}{\boldsymbol{z}_{23}^{2}} K_0 (QZ) \right) - \left( 1 \leftrightarrow 2 \right) \right] \nonumber \\ 
    - \chi_{\gamma^+ \gamma^5 \sigma} \left[ \left( 4 \epsilon^{ \sigma \rho + - } \frac{x_1 x_2}{1-x_2} \frac{Q}{Z} K_1 (QZ) +  T_2^{\sigma \rho \nu} (\{ x_i \})  \frac{z_{23 \perp \nu}}{\boldsymbol{z}_{23}^{2}} K_0 (QZ) \right) + \left( 1 \leftrightarrow 2 \right) \right] \bigg \} \; ,
\label{Eq:GeneralPhotonMesonWaveFunOv}
\end{gather}
where $c_f = N_c^2/(N_c^2-1)$, the $\chi$'s functions are Fourier transforms of the 3-body vacuum-to-meson matrix elements, i.e.
\begin{gather}
    \chi_{\Gamma^{\lambda} , \sigma } 
   = \hspace{-0.1 cm} \int_{-\infty}^{\infty} \hspace{-0.1 cm} \frac{ {\rm d} z_{1}^{-}}{2 \pi} \frac{ {\rm d} z_{2}^{-}}{2 \pi} \frac{ {\rm d} z_{3}^{-}}{2 \pi} {\rm e}^{-ix_{1}q^{+}z_{1}^{-}-ix_{2}q^{+}z_{2}^{-}-ix_{3}q^{+}z_{3}^{-}}  \left\langle M\left(p_{M}\right)\left|\overline{\psi}\left(z_{1}\right) g \Gamma^{\lambda} F_{-\sigma}\left(z_{3}\right)\psi\left(z_{2}\right)\right|0\right\rangle _{z_{1,2,3}^{+}=0} ,
   \label{Eq:Chi+_Vector_function}
\end{gather}
and the tensor structures, $T_i^{\sigma \rho \nu}$, are given by
\begin{gather}
   T_1^{\sigma \rho \nu} (\{ x_i \}) = 4 \left[ 2 \frac{x_1 (\bar{x}_1 + x_2)}{x_3} g_{\perp \perp}^{\sigma \nu} q^{\rho} + \left( \frac{(\bar{x}_1 + x_2) (\bar{x}_1 - x_1)}{\bar{x}_1 x_3} g_{\perp \perp}^{\sigma \nu} g_{\perp \perp}^{\rho \mu} - \frac{1}{\bar{x}_1} \left(  g_{\perp \perp}^{\nu \rho} g_{\perp \perp}^{\sigma \mu} - g_{\perp \perp}^{\rho \sigma} g_{\perp \perp}^{\nu \mu} \right) \right) i \partial_{z_{1\perp} \mu}  \right] ,
\end{gather} 
\begin{gather}
    T_2^{\sigma \rho \nu} (\{ x_i \}) = \frac{ 4 i }{ \bar{x}_1 } \left[ 2 x_1 \bar{x}_1 q^{\rho} \epsilon^{ \nu \sigma + -} - \left( \left( 1 + \frac{2 x_2}{x_3} \right) \left( g_{\perp \perp}^{ \sigma \mu } \epsilon^{\nu \rho + -} -  g_{\perp \perp}^{ \rho \sigma } \epsilon^{\nu \mu + -}  \right)  + (x_1 - \bar{x}_1) g_{\perp \perp}^{ \rho \mu }  \epsilon^{\nu \sigma + -} \right ) i \partial_{z_{1\perp} \mu}  \right ] \; ,
\end{gather}
\end{widetext}
with $\bar{x}_i = 1 - x_i$. In eq.~(\ref{Eq:GeneralPhotonMesonWaveFunOv}), the $K_{\nu}$'s are MacDonald functions, $Z = \sqrt{x_{1}x_{2}\boldsymbol{z}_{12}^{2}+x_{1}x_{3}\boldsymbol{z}_{13}^{2}+x_{2}x_{3}\boldsymbol{z}_{23}^{2}}$ and the $(1 \leftrightarrow 2)$ symbol denotes the following exchange: $\left( x_1 , \boldsymbol{z}_1 \right) \leftrightarrow \left( x_2 , \boldsymbol{z}_2 \right)$.
\noindent We emphasize again that the 3-body PMOW in eq.~(\ref{Eq:GeneralPhotonMesonWaveFunOv}) contains infinite kinematic twist effects through the vacuum-to-meson matrix elements. 

\section*{Expansion to twist 3}
In this section, we perform a kinematic twist expansion of the results in eqs.~(\ref{Eq:StandardDipoleAmp-rb}, \ref{Eq:Psi_2_Full_KinTwist}) and eqs.~(\ref{Eq:GenealStructure3bodyWaveFunOver}, \ref{Eq:GeneralPhotonMesonWaveFunOv}), in terms of twist-3 distribution amplitudes (DAs)~\cite{Ball:1998sk}. The expansion in powers of the hard scale is governed by contributions from small transverse separations between the constituents. 
The expansion of a non-local correlator (e.g. the one in~(\ref{Phi_+_Distrib_Coordinate})) in powers of $r^2$ is the subject of the operator product expansion (OPE) à la Balitsky-Braun~\cite{Balitsky:1987bk}, in which this latter is expanded in powers of the deviation from the light-cone $r^2  = 0$.
Each coefficient of this expansion is a finite sum of on-light-cone non-local correlators\footnote{This is different from the more standard Wilson OPE, in which each coefficient of the $r^2$ expansion is a series of local operators.}. In our case, the leading term in this expansion is represented by the initial correlator evaluated for light-like separation, while the subleading contributions to the OPE are at least twist-4~\cite{Ball:1998sk} and can be neglected within our accuracy. Although within twist-3 accuracy, the non-local correlator in eq.~(\ref{Phi_+_Distrib_Coordinate}) is taken on the light-cone, the parametrization of the vacuum-to-meson non-perturbative matrix element leads to contributions of different kinematic twist. The reason is that, the matrix element can only be a linear combination of the available four-vectors: $p_{M \mu}, r_{\mu}$ and $\varepsilon_{M \mu}^{*}$ (the latter is the meson polarization vector), with some coefficients depending on the available Lorentz invariants, $p_M \cdot r, \varepsilon_M \cdot r$ and $m_M^2$\footnote{We stress that at this point $r^2=0$.}. These quantities have different scaling in the $ Q \rightarrow \infty$ limit and this leads to contributions of different  kinematic twist.  
For example, limiting again to twist-3 accuracy, the parametrization of the 2-body vector matrix element reads~\cite{Ball:1998sk}
\begin{gather}
    \left\langle M(p_M)\left| \overline{\psi}\left(r\right)\gamma^{\mu}[r,0]\psi\left(0\right)\right|0\right\rangle \big|_{r^2 = 0} \nonumber \\ \sim f_{M}m_{M} \hspace{-0.15 cm} \int_{0}^{1} \hspace{-0.15 cm} {\rm d}x {\rm e}^{ix(p_M \cdot r)}\left[ p_M^{\mu}\frac{(\varepsilon_M^{\ast}\cdot r)}{(p_M \cdot r)}\phi(x) + \varepsilon_{M, T}^{\ast\mu}g_{\perp}^{(v)}(x) \right] ,
    \label{Eq:Twist_expa_example}
\end{gather}
where $\varepsilon_{M, T}^{\ast\lambda}$ is the transverse part of the meson polarization vector in the $(p,z)$  Sudakov basis and $\phi (x)$ and $g_{\perp}^{(v)} (x)$ are respectively twist-2 and twist-3 DAs. A convenient way of understanding the twist counting is to move to the infinite momentum frame in which $p_M^+$ and $r^{-}$ are the only non-zero components of $p_M$ and $r$. Then, since $p_M^+ \sim Q \rightarrow \infty$, $\varepsilon_{M, T}^{*} \sim 1$ and $(p_M \cdot r) \sim (\varepsilon_{M,T}^{*} \cdot r) \sim 1$, this implies that the first term in eq.~(\ref{Eq:Twist_expa_example}) scales as $Q$, while the second scales as 1. \\

A subtlety, in the present case, is due to the fact that the correlators appearing in eqs.~(\ref{Phi_+_Distrib_Coordinate}) and (\ref{Eq:Chi+_Vector_function}) are not expressed in the gauge invariant form of ref.~\cite{Balitsky:1987bk,Ball:1998sk}, i.e. they do not contain gauge links between the various fields. Nevertheless, the parameterization for correlators without gauge links can be obtained, within the required twist approximation, by expanding the Wilson lines inside the gauge invariant correlators. Indeed, in the expansion of the gauge links, e.g.  
\begin{gather}
     \mathcal{P} \, {\rm exp} \left[ i g \hspace{-0.05 cm} \int_0^1 \hspace{-0.2 cm} dt A^{\mu} ( tz ) z_{\mu} \right] \hspace{-0.1 cm} = \hspace{-0.05 cm} 1 \hspace{-0.05 cm}+ \hspace{-0.05 cm} ig \int_0^1 \hspace{-0.2 cm} dt  A^{\mu} \left( tz \right) z_{\mu} + ... \; ,
    \label{Eq:GaugeLink}
\end{gather}
the twist increases with the expansion order.
This implies that in the 3-body contribution the gauge link effect is twist-4, and can be neglected. Then, we obtain the 3-body twist-expanded PMWO
\begin{widetext}
\begin{gather}
    \Psi_3 \left( x_1, x_2, x_3 , \boldsymbol{z}_1, \boldsymbol{z}_2, \boldsymbol{z}_3 \right) = \frac{e_{q} m_M c_f}{8 \pi } \delta \left(1 - \frac{p_M^+}{q^+} \right) \left(\varepsilon_{q\rho}-\frac{\varepsilon_{q}^{+}}{q^{+}}q_{\rho}\right)  \; \left( \varepsilon^{* \mu }_{M} - \frac{p_{M}^{\mu}}{p_M^+} \varepsilon_{M}^{*+} \right) \left( \prod_{j=1}^{3} \theta (x_j) \theta (1-x_j) e^{-i x_j \boldsymbol{p}_M \boldsymbol{z}_j } \right)   \nonumber \\
    \times  \bigg \{ - i f_{3M}^{V} g_{\sigma \mu} V(x_1, x_2) \left[ \left( 4 i g_{\perp \perp}^{ \rho \sigma } \frac{x_1 x_2}{1-x_2} \frac{Q}{Z} K_1 (QZ) + T_1^{\sigma \rho \nu} ( \{ x_i \} ) \frac{z_{23 \perp \nu}}{\boldsymbol{z}_{23}^{2}} K_0 (QZ) \right) - \left( 1 \leftrightarrow 2 \right) \right] \nonumber \\ 
    - \epsilon_{- + \sigma \beta} f_{3M}^{A} g^{\beta}_{\perp \perp \mu} A (x_1, x_2) \left[ \left( 4 \epsilon^{ \sigma \rho + - } \frac{x_1 x_2}{1-x_2} \frac{Q}{Z} K_1 (QZ) +  T_2^{\sigma \rho \nu} ( \{ x_i \} ) \frac{z_{23 \perp \nu}}{\boldsymbol{z}_{23}^{2}} K_0 (QZ) \right) + \left( 1 \leftrightarrow 2 \right) \right] \bigg \} \; ,
    \label{Eq:PMOW_3Body_Twist-3}
\end{gather}
where $m_M$ is the meson mass. The PMWO in eq.~(\ref{Eq:PMOW_3Body_Twist-3}) is expressed in terms of the genuine twist-3 collinear vector, $V(x_1, x_2)$, and axial, $A(x_1, x_2)$, DAs and $f_M^V$ and $f_M^A$ are the corresponding normalization constants~\cite{Ball:1998sk}. \\

In the 2-body contribution, eq.~(\ref{Eq:Psi_2_Full_KinTwist}), in order to express the correlators in~(\ref{Phi_+_Distrib_Coordinate}), in terms of the gauge invariant ones, the effect of the second term in the expansion~(\ref{Eq:GaugeLink}) is twist-3 and must be taken into account. Since this term introduces an additional transverse gluon, it can be parameterized in terms of the genuine twist-3 DAs introduced above. Then, we obtain the 2-body twist-expanded PMWO in the form 
\begin{gather}
     \Psi_2 \left(x, \boldsymbol{r} \right) = e_q m_M f_M \delta \left( 1 - \frac{p_M^+}{q^+} \right) \left(\varepsilon_{q\mu}-\frac{\varepsilon_{q}^{+}}{q^{+}}q_{\mu} \right) \left(\varepsilon_{M \alpha}^{*}-\frac{\varepsilon_{M}^{*+}}{p_M^{+}} p_{M \alpha} \right) \nonumber \\ \times \left[ -i r_{\perp}^{\alpha} (h(x) - \tilde{h}(x)) \left( 2 x \bar{x} q^{\mu} + (x-\bar{x}) \frac{- i \partial}{ \partial r_{\perp \mu}} \right) + \epsilon^{\mu \nu + -} \epsilon^{+ \alpha - \delta} r_{\perp \delta} \left( \frac{g^{(a)}_{\perp} (x) - \tilde{g}_{\perp}^{(a)} (x)}{4} \right) \frac{\partial}{\partial r_{\perp}^{\nu}} \right] K_0 \left( \sqrt{x \bar{x} Q^2 \boldsymbol{r}^{2}} \right) ,
     \label{Eq:PMOW_2Body_Twist-3}
\end{gather}
which depends on the independent kinematic twist-3 DAs $h(x)$\footnote{Please note that $h(x)$ and $g_{\perp}^{(v)}$ are related through the relation $h(x)=\int_{0}^{x}{\rm d}u\left(\phi(u)-g_{\perp}^{(v)}(u)\right)$.} and $g^{(a)}_{\perp} (x)$ and on $\tilde{h}(x)$ and $\tilde{g}_{\perp}^{(a)} (x)$  gauge-link related terms, given by
\begin{gather}
  \widetilde{h}\left(x\right)=\frac{f_{3M}^{V}}{f_{M}}\int_{0}^{x}{\rm d}x_{q}\int_{0}^{1-x}\!{\rm d}x_{\overline{q}}\,\frac{V\left(x_{q},x_{\overline{q}}\right)}{\left(1-x_{q}-x_{\overline{q}}\right)^{2}} \; , \hspace{1 cm} \widetilde{g}_{\perp}^{\left(a\right)}\left(x\right)=4\frac{f_{3M}^{A}}{f_{M}}\int_{0}^{x}{\rm d}x_{q}\int_{0}^{1-x}\!{\rm d}x_{\overline{q}}\,\frac{A\left(x_{q},x_{\overline{q}}\right)}{(1-x_{q}-x_{\overline{q}}+i\epsilon)^{2}} \; .  
\end{gather}
The $f_M$ appearing in eq.~(\ref{Eq:PMOW_2Body_Twist-3}) is the standard vector decay constant~\cite{Ball:1998sk}.
\end{widetext}

\section*{Conclusion}
We have considered exclusive diffractive light vector-meson production with both incoming photon and outgoing meson carrying arbitrary polarizations and in the fully general non-forward kinematics. Our main results are contained in eq.~(\ref{Eq:PMOW_3Body_Twist-3}) and~(\ref{Eq:PMOW_2Body_Twist-3}), where we expressed the PMWO in terms of twist-3 DAs. We performed a consistency check of our calculations by reproducing the results of refs.~\cite{Anikin:2009hk,Anikin:2009bf}, valid in the forward and dilute limit, where only the transverse photon to transverse meson transition survives (see ref.~\cite{Boussarie:2024bdo} for details). \\ 

These results have a series of implications. First of all, they provide the most complete description of the DVMP processes in the high-energy regime, offering many future phenomenological applications. Second, they enlarge the list of processes useful to probe the gluon GTMD distribution. Lastly, but more importantly, such a general description has required a systematic higher-twist treatment of exclusive processes in the saturation framework, in order to describe helicity amplitudes entering the process at the subleading power in $Q$. Our general method paves the way for higher-twist studies at small-$x$ and, in particular, in the saturation regime where hard scale power-suppressed contributions are important in hunting for precision. Besides, there is a dual utility in considering higher-twist effects in the context of $k_T$-factorization. In fact, the applicability of collinear factorization at the twist-3 accuracy is far from obvious and, in some cases like the one discussed in this letter, it leads to end-point singularities that make any theoretical prediction problematic. In such situations, a (generalized) $k_T$-dependent factorization, reliable at small-$x$, is a valid option, since the end-point singularities are naturally regularized by the transverse momenta of the $t$-channel gluons.

\acknowledgments
We thank Valerio Bertone, Guillame Beuf, Giovanni A. Chirilli, Andrey V. Grabovsky, Edmond Iancu, Saad Nabeebaccus, Alessandro Papa, Simone Rodini and Jakob Schoenleber for useful discussions. This  project  has  received  funding  from  the  European  Union’s  Horizon  2020  research  and  innovation program under grant agreement STRONG–2020 (WP 13 "NA-Small-x"). The work by M.~F. is supported by Agence Nationale de la Recherche under the contract ANR-17-CE31-0019. M. F. acknowledges support from the Italian Foundation “Angelo Della Riccia”. The  work of  L.~S. is  supported  by  the  grant  2019/33/B/ST2/02588  of  the  National  Science Center in  Poland. L.~S. thanks the P2IO Laboratory of Excellence (Programme Investissements d'Avenir ANR-10-LABEX-0038) and the P2I - Graduate School of Physics of Paris-Saclay University for support. This work was also partly supported by the French CNRS via the GDR QCD.

\end{document}